\documentclass[conference, a4paper]{IEEEtran}
\IEEEoverridecommandlockouts

\usepackage{soul}
\usepackage[bookmarksopen=true]{hyperref}
\usepackage{fancyhdr}
\usepackage{cite}
\usepackage{graphicx}
\usepackage{psfrag}
\usepackage{subfigure}
\usepackage{url}
\usepackage{balance}
\usepackage{stfloats}
\usepackage{amsmath,amsthm,amssymb}

\usepackage{array}
\usepackage{fancyhdr}
\usepackage{float}
\usepackage{epsfig}
\usepackage{color}
\usepackage[nomain,acronym,toc]{glossaries}

\usepackage{xcolor}
\usepackage[linesnumbered,ruled,vlined]{algorithm2e}

\theoremstyle{definition}

\definecolor{sblue}{RGB}{0,51,120}
\definecolor{sred}{RGB}{200,51,130}

\SetKwInput{KwInput}{Input}                
\SetKwInput{KwOutput}{Output}              

\def\BibTeX{{\rm B\kern-.05em{\sc i\kern-.025em b}\kern-.08em
		T\kern-.1667em\lower.7ex\hbox{E}\kern-.125emX}}
\begin{document}
	
	\title{Correlation-Based Device Energy-Efficient Dynamic Multi-Task Offloading for Mobile Edge Computing 
	}
	
	\author{\IEEEauthorblockN{Siqi Zhang, Na Yi and Yi Ma\\}	
		\IEEEauthorblockA{Institute for Communication Systems, University of Surrey, UK, GU2 7XH\\
			 e-mails: \{s.zhang, n.yi, y.ma\}@surrey.ac.uk}
	}

	\maketitle

\begin{abstract}	  Task offloading to mobile edge computing (MEC) has emerged as a key technology to alleviate the computation workloads of mobile devices and decrease service latency for the computation-intensive applications. Device battery consumption is one of the limiting factors needs to be considered during task offloading. In this paper, multi-task offloading strategies have been investigated to improve device energy efficiency. Correlations among tasks in time domain as well as task domain are proposed to be employed to reduce the number of tasks to be transmitted to MEC. Furthermore, a binary decision tree based algorithm is investigated to jointly optimize the mobile device clock frequency, transmission power, structure and number of tasks to be transmitted. MATLAB based simulation is employed to demonstrate the performance of our proposed algorithm. It is observed that the proposed dynamic multi-task offloading strategies can reduce the total energy consumption at device along various transmit power versus noise power point compared with the conventional one. 

\end{abstract}	
\section{Introduction}
With the continuous development of mobile communication technology and the rapid development of mobile Internet, mobile terminals represented by smart phones, tablet computers, laptop, and smart assistants have been widely used. But the mobile terminal receives limiting factors such as volume, weight, performance, power, etc. Its working ability is still in a serious and tedious state, which cannot meet the increasing demand of people. Although the mobile terminal has made great progress in hardware technology (for example, the continuous replacement of CPU/GPU, the continuous improvement of chip manufacturing process from 28nm to 14nm to the current 7nm, 5nm\cite{8776580}, etc.), but it is still far from what people need. Moreover, with the emergence of new concepts such as autonomous driving, telemedicine, and Industry 4.0 which need ultra reliability, low latency\cite{7847322}, ordinary equipment is even more unable to support their operations. Meanwhile, with the emergence of machine learning, artificial intelligence and other emerging technologies\cite{7951770}, the rapid development of image recognition, speech recognition and other applications, virtual reality and augmented reality game applications are emerging in endlessly. The operation of these applications requires a large amount of computing resources and storage resources, and they are all computationally intensive applications at a time. Due to the limitation of  mobile terminals or some other devices, when computationally intensive applications\cite{7951770}  are running on smart terminals, the endurance of the terminal and the performance of the application are very problematic. How to solve this problem of resource limitation and energy consumption has become a huge challenge today. 

Most of the literature suggests to change the task allocation method, transmission power, clock frequency to optimize the task offloading algorithm. In \cite{8279411}--\cite{8638800}, authors proposed task splitting, which is a way to change the  task structure to reduce the latency and improve the local device energy efficiency. However, they only split the task, but did not consider the redundancy of sources. In this paper,  we propose to split task to the smallest executable task, named as unit, and then select the unit by using the correlation between them. Furthermore, both time and task domain correlation is considered in our work to selected the necessary tasks to be processed.

\section{Problem Formulation and System Models}
In this section, the definition of task and unit will be introduced first to help to understand the proposed models. Both task and unit are the process of collecting data, processing data, and sending instructions, but unit is the smallest section that can form a task, and unit cannot be divided anymore. That means task can be composed of one or more units, and a task can be split into one or more units.

Consider $N$  devices (users), denoted by a set of  $\mathbb{N}=\{1, 2, ...., N\}$, and device $i$ has $M_i$ tasks at the same time,  $i\in \mathbb{N}$, denoted by a set of $\mathbb{M}=\{M_1, M_2, ...., M_N\}$, each task in the $M_i$ tasks is composed of $M_s$ different unit, denoted by a set of $\mathbb{K}=\{k_{M_1},k_ {M_2}, ...., k_{M_s}\}$and $i\in \mathbb{N}$. After some tasks are split, some identical units may be generated, so the correlation of task domain came into being. In our work, it is proposed to use the correlation between units to improve the energy-efficient of local device and reduce the latency. Moreover, it is assumed here that devices communicate with MEC server orthogonally. The main target here is to improve the energy efficiency of local devices when fulfil the extremely severe latency requirements of each unit and each device.

The energy cost  minimization problem is formulated as:
\begin{equation}\label{eqn01}
OPT-1\ \ \ \ \ \ \ \ \ \min \limits_{\mathcal{A,\ F,\ P}}\sum_{n=1}^{n=N}E_n^L
\end{equation} 
Subject to{\color{black}{
\begin{equation}
		\begin{split}
C1: LT_{n,j}^m&\leq T_{n,j\ max},\ \ \ \ \ j \in A, n\in \mathbb{N}\\
C2: LT_{n,k}^l&\leq T_{n,k\ max},\ \ \ \ \ k \in B, n\in \mathbb{N}\\
C3: LT_{n}&\leq T_{n\ max},\ \ \ \ \  n\in\mathbb{N}\\
C4:f_{n,l}&\leq f_{n,l\ max},\ \ \ \ \ \ n\in \mathbb{N}\\
C5:p^t_{n}&\leq p^t_{n\ max}, \ \ \ \ \ \ n\in \mathbb{N}
		\end{split}
	\end{equation}}}

$E_n^L$ represents the local energy consumption of user $n$, $ LT_{n,j}^m$ represents the latency of unit $j$ of user n processed on MEC,  $LT_{n,k}^l$ represents the latency of unit k of user $n$ processed on local device, $ T_{n,j\ max}$ represents the latency requirement of unit $j$,  $A$ represents the collection of all tasks processed on the MEC, and $B$ represents the collection of all units processed on the local device, $LT_n$  represents the latency of user $n$,  $T_{n\ max}$ represents the latency requirement  of user $n$,  $f_{n,l}$   represents the computation capability of user $n$, such as the number of CPU cycles per second,  $ p^t_{n}$ represents the transmission power of  user $n$, and $f_{n,l\ max},\  p^t_{n\ max}$ indicate the maximum value of $f_{n,l}$ and $ p^t_{n}$ respectively.

 $\mathcal{F}=\{f_{n,l}|n\in \mathbb{N}\}$,  $\mathcal{P}=\{p^t_{n}|n\in \mathbb{N}\}$, $\mathcal{A}= \{A_{n,j}|n\in \mathbb{N},j \in \mathbb{M}2\}$, C1 and C2 are to limit the latency of each task to not exceed the requirements, C3  is to limit the user's overall latency does not exceed the requirements, C4 is to limit the processing power of the local device not to exceed its maximum processing power, C5 is to limit the transmission power of the local device not to exceed its maximum transmission power.
 
\subsection{Transmission Model}
{\color{black}{When the interference is not considered}}, the signal-to-noise ratio (SNR) of user n is
\begin{equation}\label{eqn07}
SNR_n=\frac{p_n^th_{n,m}^2}{\color{black}{B_w\mathcal{N}_0}}
\end{equation} 
and then, the transmission rate (uplink) of user $n$ is calculated as:
\begin{equation}\label{eqn08}
r_n=Blog_2(1+ SNR_n).
\end{equation} 
In \eqref{eqn07} and \eqref{eqn08}, {\color{black}{$B_w$}} represents the bandwidth of this channel,  $h_{n, m}$  represent the channel gain of user n   to MEC, and {\color{black}{$\mathcal{N}_0$ represents the noise spectral density}}, $p_n^t$ represents the transmission power of user $n$. Because the amount of data that needs to be offloaded is much larger than the amount of data that needs to be downloaded, {\color{black}{so the downlink is ignored in this paper}}\cite{7524497,7553459}.

\subsection{Computation Model}
Let $w_{n,j}$ {\color{black}{represent}} the CPU cycle required to calculate the unit (unit $j$ of user $n$), $d_{n,j}$ represents the computation input data (in bits) of the {\color{black}{ $j$ th unit}} of {\color{black}{$n$ th user of local devices}}.
\subsubsection*{Local Computing}
{\color{black}{when user chooses to process locally, use}} $f_{n,l}$ to represent the {\color{black}{CPU clock speed}} of user's device, so the latency in local for computing can be represented as: 
\begin{equation}\label{eqn09}
t_{n,j}^l=\frac{w_{n,j}}{f_{n,l}}.
\end{equation} 
where $ t_{n,j}^l$ {\color{black}{denotes}} the time required to complete unit $j$ of user $n$ in local device.

The energy required to complete unit $j$ in local device can be expressed by the following formula
\begin{equation}\label{eqn10}
E_{n,j}^l=\kappa w_{n,j}f^2_{n,l}
\end{equation} 
In this case, $\kappa$ is the effective switched capacitance depending on the chip architecture\cite{miettinen2010energy}.
\subsubsection{MEC Computing}
when a user {\color{black}{decides}} to offload tasks to MEC, and {\color{black}{transmit}} a unit through a wireless network, the corresponding transmission latency and energy consumption will be generated. According to the communication model, the uplink transmission latency when user $n$ offload the task is:
\begin{equation}\label{eqn11}
t_{n,j}^t=\frac{d_{n,j}}{r_n},
\end{equation} 
and the energy required to transmit unit $j$ can be expressed by the following formula:
\begin{equation}\label{eqn12}
E_{n,j}^{t}=P_{n,j}^{t}t_{n,j}^t.
\end{equation} 
{\color{black}{where}} $ t_{n,j}^t$ represents the time required to transmit the {\color{black}{$j$ th unit of $n$ th user.}}
After the computing unit is offloaded to the MEC, the MEC will allocate certain computing resources to this unit, considering that the computing resources allocated by the MEC to each user are fixed. Let $f_{MEC}$ denote the computing resources allocated by the MEC, and the latency for the MEC to perform  {\color{black}{the $j$ th unit of $n$ th user}} is
\begin{equation}\label{eqn13}
t_{n,j}^m=\frac{w_{n,j}}{f_{MEC}}.
\end{equation} 
Because MEC has a constant energy supply, {\color{black}{MEC's energy consumption need not be considered in this paper.}}

\begin{figure}[t]
\centering
\includegraphics[width=8cm,height=3cm]{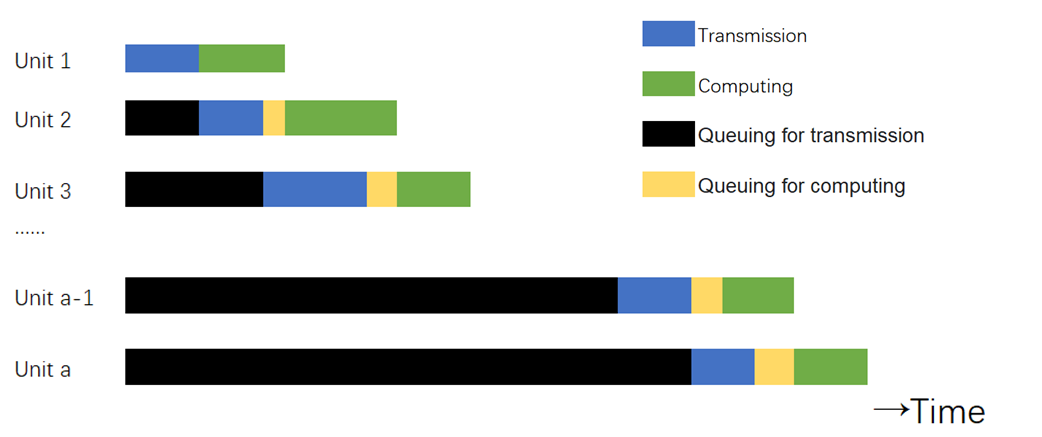}
\vspace{-1.5em}\caption{{\color{black}{Tasks which are processed in MEC}}}
\label{figure1}
\end{figure}

\begin{figure}[t]
\centering
\vspace{-1em}\includegraphics[width=8cm,height=3cm]{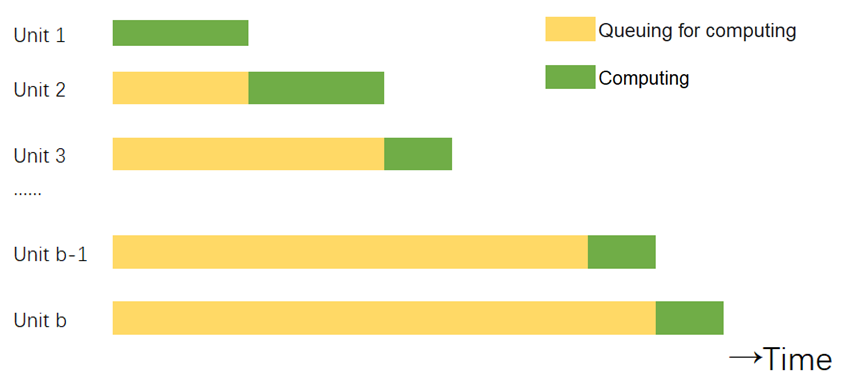}
\vspace{-1em}\caption{{\color{black}{Tasks which are processed in local}}}
\label{figure2}
\end{figure}

In this paper, {\color{black}{it is assumed}} that  MEC can only process {\color{black}{one}} unit at the same time, communication capacity can only support the transmission of a unit simultaneously, as shown in Fig. \ref{figure1} and Fig. \ref{figure2}. Blue {\color{black}{blocks}} represent transmission time, green {\color{black}{ones}} represent computing time, black ones {\color{black}{and yellow ones represent queuing for transmission, and queuing for computing respectively}}
At the beginning of this section, {\color{black}{it is introduced}} that each user generates multiple tasks and then they are split into multiple units, and needs to be processed. However, due to channel and processor limitations, these units cannot be processed at the same time. {\color{black}{There are}} two definitions of waiting time for the units which will be processed in local and waiting time for the units which will be processed in MEC. Suppose A is the set of units which will be processed in MEC, and A=\{1,2, ...,a\}, B is the set of units which will be processed in local device, and B=\{1,2, ...,b\}, 
\par \textbf{Definition 1}  (waiting time for the units which will be processed in local): The sum of all time, before the unit is processed by local device. {\color{black}{Use}} $WT_{n,k}^l$  to present the waiting time of unit $k$ of device $n$, which will be processed in local.
\par \textbf{Definition 2} (waiting time for the units which will be processed in MEC): The sum of all time, before the unit is processed in MEC. {\color{black}{Use}} $WT_{n,j}^m$  to present the waiting time of unit $j$ of device $n$, which will be processed in MEC.

According to {\color{black}{the}} Definition 2, waiting time for the units which will be processed in MEC {\color{black}{is divided}} into {\color{black}{two}} parts, time of  waiting for transmission plus time of transmission ($WT3_{n,j}^m$), and 
time before processing in MEC and after MEC receives the unit($WT4_{n,j}^m$), that means $ WT_{n,j}^m$ {\color{black}{can be divided}} into $WT3_{n,j}^m+ WT4_{n,j}^m$.

Assume that $unit_j\in A$\\
$when\ j=1$
\begin{equation}\label{eqn14}
WT3_{n,j}^m=t_{n,j}^t,\ \ \ \ \ WT4_{n,j}^m=0
\end{equation}
$when\ j >1$
{\color{black}{
\begin{equation}\label{eqn15}
\begin{split}
WT3_{n,j}^m&=WT3_{n,j-1}^m+t_{n,j}^t\\
WT4_{n,j}^m&=max(WT3_{n,j}^m, LT_{n,j-1}^m)-WT3_{n,j}^m
\end{split}
\end{equation} 
In (\ref{eqn15}) , max means the maximum value between these two value.
\begin{equation}\label{eqn18}
\begin{split}
WT_{n,j}^m&=WT3_{n,j}^m+WT4_{n,j}^m\\
LT_{n,j}^m&=WT_{n,j}+t_{n,j}^m
\end{split}
\end{equation}
$LT_{n,j}^m$ means the latency of $unit_j$, $unit_j\in A$

Assume that $unit_k\in B$\\
so, in this case
\begin{equation}\label{eqn20}
WT_{n,k}^l=\sum_{c=1}^{c=j-1}t_{n,c}^l
\end{equation} 
so that, the latency of unit $k$ can be expressed as ($unit_k\in B$):
\begin{equation}\label{eqn21}
LT_{n,k}^l=\sum_{c=1}^{c=k-1}t_{n,c}^l+t_{n,k}^l=\sum_{c=1}^{c=k}t_{n,c}^l
\end{equation} 
So the latency of the whole system of user $n$ can be represent as:
\begin{equation}\label{eqn22}
Ts_n=max(LT_{n,b}^l, LT_{n,a}^m)
\end{equation} 
Use $E_n^L$ to represent the total energy consumption of local device $n$:
\begin{equation}\label{eqn23}
E_n^L=\sum_{a=1}^{a=j} E^t_{n,a}+\sum_{b=1}^{b=k} E^l_{n,b}
\end{equation} 
}}
\section{Proposed task offloading algorithm}
In this {\color{black}{section}}, first, several preprocessing methods {\color{black}{will be proposed}}, and perform some operations before task offloading, without affecting the reliability of the task, reduce the amount of task calculation and transmission, and then introduce the algorithm proposed for task offloading. 
\subsection{Correlation in Time Domain of Tasks}
With the change of time, the task constantly updates its own information source to process the task. In the traditional task offloading, only consider how to change the task allocation mode, the processing frequency, transmission power to reduce latency and improve the energy-efficient, but in their algorithm, the data size does not change, which to a large extent, restricts the development of  task offloading, {\color{black}{this paper provide a new way}} to filter some unnecessary information to reduce the data size, so as to further reduce the latency and energy consumption. Correlation coefficient is a good way to be considered.

The correlation coefficient {\color{black} introduced in \cite{7000704}} is defined as:
\begin{equation}\label{eq1}
r(x,y) \triangleq  \frac{\mathsf{cov}(x,y)}{\sqrt{\mathsf{var}(x)\mathsf{var}(y)}}\ ,\ r(x,y) \in [-1,1].
\end{equation}
In this formula, x, y are the sources of information {\color{black}{which need to be compared}}. Suppose there is a task, denoted as $Task\ A$, {\color{black}{and}} denote the names of $Task\ A$ at different moments as $Task\ A1$, $Task\ A2$, $Task\ A3$, $Task\ A4$ in the order of time, $A1$ is the first and $A4$ is the last one, to reduce the number of processing, {\color{black}{correlation coefficient should be used}} between  them to decide whether to process these tasks, or process part of them. There are two ways to use the correlation coefficient.

The first way, set only one threshold to decide whether to process this task:  set a threshold $\alpha$, and then process the $Task\ A1$, and then calculate then correlation coefficient of $Task\ A1$ and $Task\ A2$. If the correlation coefficient between them is greater than $\alpha$, the $Task\ A2$ {\color{black}{will not be processed}}, and keep the $Task\ A1$ in memory, and the correlation coefficient between $Task\ A1$ and $Task\ A3$ {\color{black}{should be calculated}}, if it is greater than the  $\alpha$, $Task\ A3$ {\color{black}{will not be processed}} too, and continue to calculate the correlation coefficient between $Taks\ A1$ and $Task\ A4$, if not, $Task\ A3$ {\color{black}{will be processed}}, and keep $Task\ A3$ in memory, and then calculate the correlation coefficient between $Task\ A3$ and $Task\ A4$, and repeat with following Tasks.

The second way, set multi-threshold to decide how to process this task: {\color{black}{two thresholds}} as an example. First of all, set two thresholds, $\alpha$ and $\beta$, and $\alpha >\beta$, and  process the first task. When the correlation coefficient {\color{black}{between}} two adjacent tasks is greater than $\beta$ and smaller than $\alpha$, {\color{black}{just}} need to process the different part of this two tasks, when the correlation coefficient {\color{black}{between}} two tasks is smaller than $\beta$, {\color{black}{it can be considered}} the information of these two tasks are totally different, so, {\color{black}{the new task need to be processed}}, when the correlation coefficient {\color{black}{between}} two tasks is greater than $\alpha$,  {\color{black}{it can be considered}} the information of these two tasks are totally same, {\color{black}{the new task need not to be processed again}}
\subsection{Correlation in Task Domain of Tasks}
The correlation coefficient {\color{black}{ is considered in}} time domain, so as to reduce the amount of data, then the next step is to do the task splitting to get the units of these necessary task. {\color{black}{As said before}}, different tasks may split into the same unit, so the correlation of task domain came into being. How to use these correlation to further optimize{ \color{black}{the}} algorithm is the main content of this part, and {\color{black}{use}} $C_{o,p}$ to represent the correlation  between unit o and unit p. $C_{o,p}\in \{0, 0.5, 1\}$. when $C_{o,p}=0$,  that means that there is no relationship between unit o and unit p, both units need to be processed, when $C_{o,p}=1$,  that means unit o and unit p are total same, {\color{black}{and}} only need to process one of them and share the result, when $C_{o,p}=0.5$, that means  unit o and unit p are different units, but use the same source information. 
\subsection{Units Allocation}
\begin{figure}[t]
\centering
\includegraphics[scale=0.27]{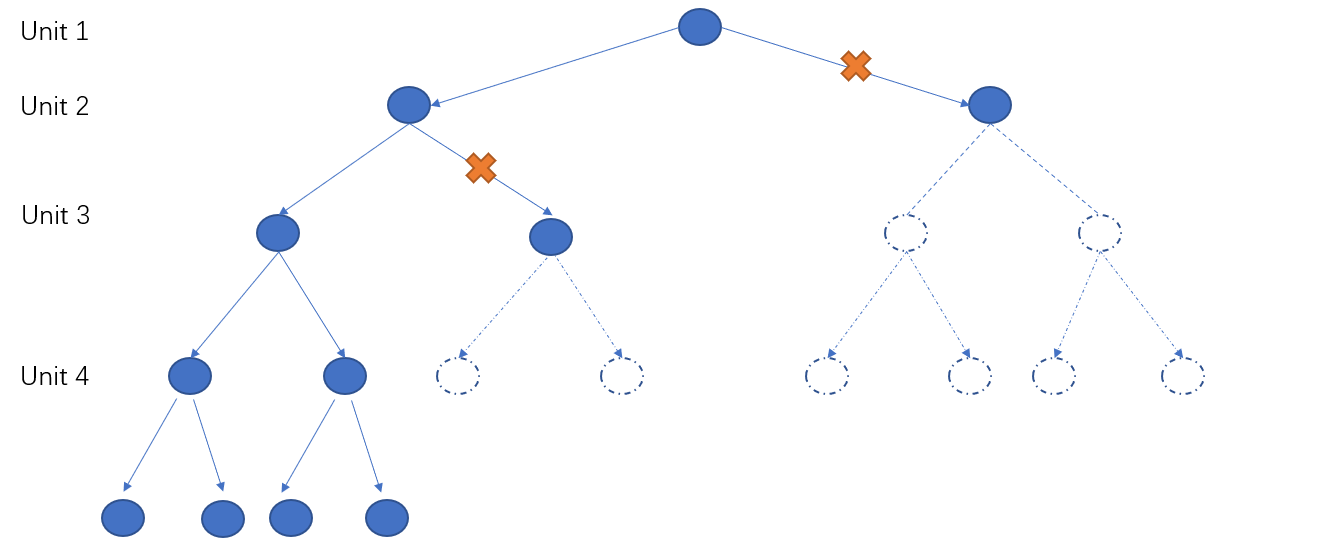}
\vspace{-1.5em}\caption{Binary Tree}
 \label{figure42}
\end{figure}

After {\color{black}{arranging units}} in the order required by the latency,  determine whether this unit needs to be assigned to the MEC for processing {\color{black}{from the first unit}}. At this time, the tree diagram{\color{black}{ needs to be drawn}}, as shown in Fig. \ref{figure42}, {\color{black}{use}}  four units as an example. Starting from the vertex first unit, there are two paths, one (unit will be processed in MEC Server) and zero (unit will be processed in local device). If the unit latency caused by the path meets the requirements of this unit, this path will be maintained and the next level of judgment will be made. If the latency does not meet the requirements, then drop this path. {\color{black}{Until finish}} the last task, and judgment whether the system latency is meet the requirement, if the answer is YES, keep this node, if not, drop this node, then all the feasible solutions {\color{black}{can be obtained}}. Then, for getting the optimal solution, some other optimization to transmission power and clock frequency {\color{black}{need to be done, and they will be introduced in the next few parts. }}
In the above situation, there is no correlation between units, {\color{black}{the next section will introduce}} the situation with there are correlation between units. when $C_{o,p}=1$,  the duplicate units {\color{black}{should be deleted}}, and then allocate the remaining units in the same way as said before. when $C_{o,p}=0.5$,  these units with a correlation degree of 0.5 {\color{black}{ need to be merged}}, that means merge them into a large task to carry out the process of task allocation(this allocation process is same as said before). If this large task is allocated to MEC server for processing, {\color{black}{it}} will reduce the latency and energy consumption caused by communication, that because they use the same source information. If this large task is allocated to local device for processing, the energy consumption and the latency will not change. 
\subsection{Clock Frequency Allocation}
According to {\color{black}{(\ref{eqn10})}}, {\color{black}{if}} $w_{n,j}$ (the CPU cycle required to calculate the unit) not change, the smaller the value of $f_n^t$, the lower energy consumption will take.
\subsection{Transmission Power Allocation}
According to {\color{black}{(\ref{eqn07}),  (\ref{eqn08}), and (\ref{eqn12})}} and use $D$ to represent the total transmission volume of all units that need to be transmitted to MEC, so the energy consumed for transmission becomes E1
\begin{equation}
E1 ={\color{black}{\frac{D}{B_w}}\times log_{\eta}}2^{p^t_n}
\end{equation}
{\color{black}{$\eta=1+\frac{p_n^th_{n,m}^2}{{B_w}\mathcal{N}_0}$.}} So, the monotonic interval of E1 with respect to $p_n^t$, and take the  $p_n^t$value that minimizes E1 under conditions C1 to C5.
\subsection{Algorithm Flow}
Use the maximum $p_n^t$ of the local device and maximum $f_n^t$ of local device to find all combinations that satisfy the conditions of C1-C5, and record them as $\mathcal{V}$, and then use the algorithms proposed before find the smallest value of energy consumption that can be achieved after optimization in $\mathcal{V}$ , take the combination with the least energy consumption as {\color{black}{the}} final result. Therefore, the optimal clock frequency, transmission power, and unit allocation method can be obtained.

\section{Simulation results}
In this section, {\color{black} we evaluate our proposed task-offloading algorithm with three state-of-the-art baselines through MATLAB based simulations. It is considered in our simulation, that system contains four users and a single MEC. Each user has different number of tasks, and different type of task has different size.} Each user has {\color{black} its dedicated orthogonal channel to communicate to MEC}.

{\color{black}{The channel bandwidth is set }}to 20MHZ, and the channel conforms to the Rayleigh distribution. The SNR of the user terminal is set to 10dB, 20dB, 30dB, 40dB and 50dB in our simulations. The maximum computing rate in user's device is  $2\times10^9$cycles/s\cite{7914660}, and the maximum computing rate in MEC server is $20\times10^9$cycles/s \cite{7914660}, the size of each task is between 1-3M, and the number of processing revolutions they require is Between 1500-4500 cycles\cite{7914660}, and $\kappa =10^{-11}$ \cite{miettinen2010energy}. {\color{black}{There are}} two latency requirements, which are 50ms and 100ms.

\begin{figure}[t]
\centering
\vspace{-1.5em}	\subfigure[SNR=20,30dB]{
		\begin{minipage}[t]{1\linewidth}
			\centering
			\includegraphics[width=0.825\textwidth]{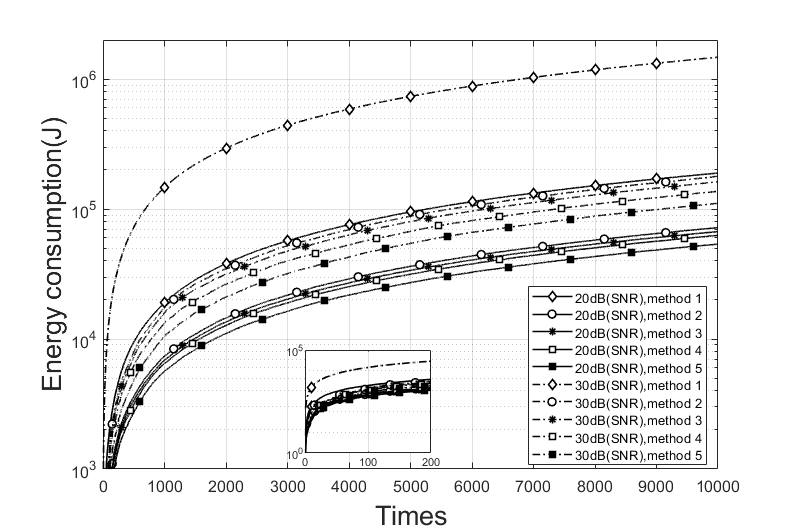}
			\label{20}
		\end{minipage}%
	}%
	
	\quad      
\subfigure[SNR=40,50dB]{
\begin{minipage}[t]{1\linewidth}
		\centering
              \includegraphics[width=0.825\textwidth]{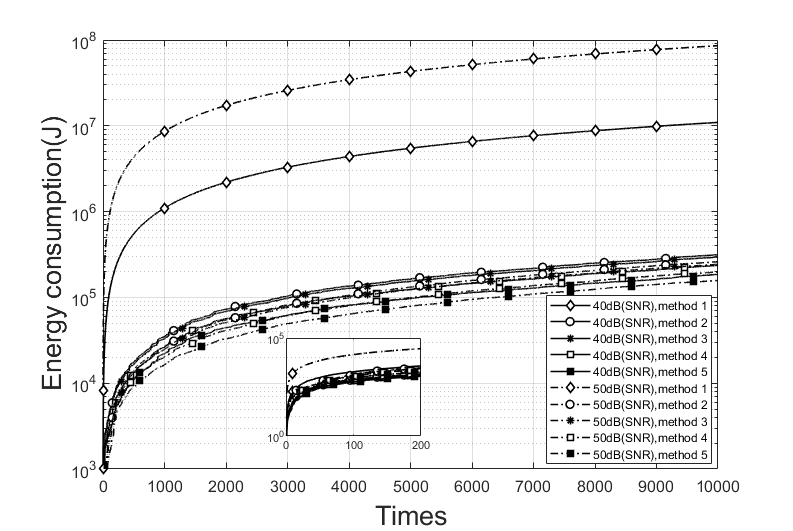}
			\label{40}
		\end{minipage}
	}%
	\centering
			\vspace{-1em}\caption{\color{black}{Energy consumption of five different algorithms }}
	\label{20304050}
\end{figure}

{\color{black}{In the experiment, {\color{black}{two proposed approaches simulated with the other three baselines}}.  Method 1 (Baseline 1): the most primitive task offloading algorithm, without considering any transmission power and other optimizations. Method 2 (Baseline 2): Further optimize the processing frequency and transmission power based on the Method 1\cite{7524497}.  Method 3 (Baseline 3): Based on Method 1 and Method 2, consider the impact of split tasks on task offloading\cite{8279411}. Method 4: Consider the relevance of task on time domain to reduce the amount of data. method 5: Base on Method 3, consider the correlation in time domain and in different tasks to reduce the amount of data.}}

\begin{figure}[t]
\centering
\includegraphics[width=9.8cm,height=6.17 cm]{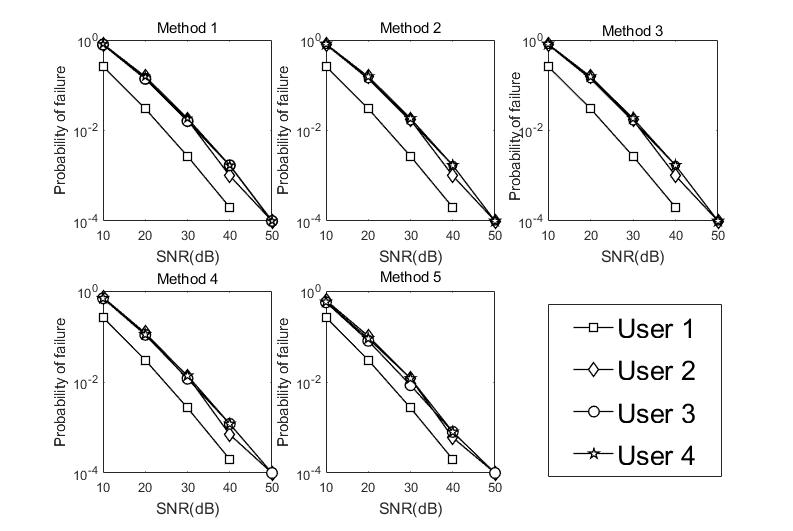}
\vspace{-1.8em}\caption{{\color{black}{ Probability of task processing failure under different methods and SNR}}}
 \label{SNRmethod}
\end{figure}

The {\color{black}{Fig. \ref{20} Fig. \ref{40} show the energy consumption of the five different algorithms on the local device  when the SNR on the user terminal are 20dB, 30dB, 40dB and 50dB respectively. In the same SNR, our proposed algorithms have better performs
 Fig. \ref{SNRmethod} shows the probability of task failure under different methods and SNR. 
From Fig. \ref{20304050}, and Fig. \ref{SNRmethod}, with the same SNR, our proposed algorithms have better performs in energy saving and reliability, and as the increase of SNR, with the same method, the energy cost may increase too, this is because we record that the energy consumption of the failed tasks that can be known in the stage of decision making is zero. The larger the SNR, the more tasks that can be processed, so it will also cause more energy consumption.}}

\section{Conclusion}
In this paper, it was studied how to improve the energy efficiency of local devices and reduce the latency in the process of task offloading. A novel task offloading algorithm has been proposed with the consideration of task correlation in both time domain and task domain to reduce the number of tasks to be transmitted. Moreover, a joint optimization has been studied for task allocation, transmission power and clock frequency. MATLAB based simulation results have demonstrated that the proposed task offloading algorithm can reduce the device energy consumption along various transmit power versus noise variance setup compare with the conventional one. 

{\color{black}{\section*{Acknowledgment}
The work was supported in part by European Com- mission under the framework of the Horizon2020 5G-Drive project, and in part by 5G Innovation Centre (5GIC) HEFEC grant.}}
\balance
	
\bibliographystyle{IEEEtran}	
\bibliography{ref}

\end{document}